\documentclass[prl]{revtex4}
\usepackage[a4paper, total={6in, 9.5in}]{geometry}     
\usepackage{graphicx,epsfig}
\usepackage{amsmath,empheq}
\usepackage{amsfonts}
\usepackage{fancyhdr}
\usepackage{float} 
\usepackage{hyperref}
\usepackage{svg}
\usepackage[export]{adjustbox}
\usepackage{bigints}
\usepackage[caption=false]{subfig}
\usepackage[export]{adjustbox}
\usepackage{balance}
\usepackage{lipsum}  
\usepackage{pdfpages} 

\begin{document}



\pagestyle{fancy}
\lhead{\bf Deformable bodies in a 3-dimensional viscous flow: Vorticity-Stream vector formulation}

\title{\textbf{Deformable bodies in a 3-dimensional viscous flow: Vorticity-Stream vector formulation}}
\author{Andreu F. Gallen$^{1,*}$} 
\author{ Joan Muñoz Biosca$^{2}$} \author{ Mario Castro$^{3,*}$} \author{Aurora Hernandez-Machado$^{2,4}$}
\affiliation{$^1$Center for Interdisciplinary Research in Biology (CIRB), Collège de France, CNRS, Inserm, Université PSL, Paris, France,}
\affiliation{$^2$Facultat de Física, Universitat de Barcelona, Diagonal
645, 08028 Barcelona, Spain,}
\affiliation{$^{3}$Instituto de Investigación Tecnológica (IIT) and Grupo Interdisciplinar de Sistemas Complejos (GISC), Universidad Pontificia Comillas, Madrid, E28015, Spain,}
\affiliation{$^{4}$Institute of Nanoscience and Nanotechnology (IN2UB), 08028 Barcelona, Spain. }
\affiliation{$^{*}$Corresponding authors: fdzgallen@gmail.com; marioc@comillas.edu}

\begin{abstract}  
When simulating three-dimensional flows interacting with deformable and elastic obstacles, current methods often encounter complexities in the governing equations and challenges in numerical implementation. In this work, we introduce a novel numerical formulation for simulating incompressible viscous flows at low Reynolds numbers in the presence of deformable interfaces. Our method employs a vorticity-stream vector formulation that significantly simplifies the fluid solver, transforming it into a set of coupled Poisson problems. The body-fluid interface is modeled using a phase field, allowing for the incorporation of various free-energy models to account for membrane bending and surface tension. 
In contrast to existing three-dimensional approaches, such as Lattice Boltzmann Methods or boundary-integral techniques, our formulation is lightweight and grounded in classical fluid mechanics principles, making it implementable with standard finite-difference techniques. We demonstrate the capabilities of our method by simulating the evolution of a single vesicle or droplet in Newtonian Poiseuille and Couette flows under different free-energy models, successfully recovering canonical axisymmetric shapes and stress profiles. Although this work primarily focuses on single-body dynamics in Newtonian suspending fluids, the framework can be extended to include body forces, inertial effects, and viscoelastic media.
\end{abstract}

\maketitle


\section{Introduction}
The study of deformable objects in microfluidic environments lies at the intersection of fluid mechanics, soft matter physics, and biomedical engineering, offering profound insights into the behavior of vesicles, droplets, and cells under confined flow conditions \cite{whitesides2006origins,baroud2010dynamics,tabeling2005introduction,fedosov2010multiscale}. In low-Reynolds number regimes typical of microfluidic systems, the interplay between hydrodynamic forces and interfacial mechanics governs the deformation, mobility, and induced resistance of these soft entities\cite{sajeesh2014hydrodynamic,sajeesh2015microfluidic,sajeesh2014particle,shen2024merged}. 
Various approaches can be employed to tackle this problem using computer simulations. One such strategy involves using a phase field coupled to a lattice Boltzmann method (LBM)~\cite{LazaroSM1,Lazaro2019}. Alternatively, particle-based methods like smoothed dissipative particle dynamics (SDPD) ~\cite{mauer2018,lanotte2016} can also be utilized. Another option is to use mathematical functions such as Green functions or Dirac deltas, for instance, through the use of boundary integral methods ~\cite{aouane2014vesicle,kaoui2011} or immersed boundary methods ~\cite{eggleton1998large}. These simulations have explored how parameters such as size, viscosity contrast, and membrane elasticity influence flow dynamics. However, implementing these methods practically can present significant challenges and computational requirements as discussed elsewhere ~\cite{aidun2010lattice,Zhao2011TheDO,Rallabandi}.  

 Recent advances, such as the vorticity–stream function formulation coupled with phase-field models, provide a robust and computationally efficient framework to simulate 2D flows interacting with elastic interfaces \cite{gallen2021red}. 
 These methods capture complex phenomena like vesicle shape transitions, lateral migration, and stress distribution, while also enabling the quantification of hydrodynamic resistance as a diagnostic tool. Such developments enhance our understanding of cellular biomechanics and droplet-based transport and pave the way for innovative lab-on-chip applications in diagnostics and targeted therapeutics. 
 {While there are other powerful fluid solving methods like Volume-of-Fluid (VOF) method, good for surface-tension problems, they lack the smooth variational framework with a differentiable interface needed to reliably derive the chemical potential and bending required for the Canham–Helfrich energy \cite{guckenberger2017theory}.}
 
In this paper, we exploit the properties of low-Reynolds flow to study the Physics of simulating elastic membranes or vesicles. Thev method generalizes the formalism introduced in Ref.~\cite{gallen2021red}  and is based on the evolution equations of three observables: the interface chemical potential, the fluid vorticity, and the so-called stream vector. The stream vector is the generalization to 3D of the widely used stream function in 2D flows, which will work as a vector potential. Importantly, adaptability is a key feature of the approach. The method allows for studying various objects with easy customization through modifying free energy, adding extra terms, or even a complete overhaul. Its coupling with obstacles that adhere to two free energy models, such as Canham-Helfrich and Cahn-Hilliard or Couette and Poiseuille, shows the method's versatility.

{ \color{black}
Existing numerical approaches for incompressible flows with deformable interfaces have primarily focused on high-Reynolds-number regimes, employing velocity–vorticity formulations and non-staggered vorticity–vector potential methods \cite{weinan1997finite,Olshanskii2010,meitz2000compact}, as well as extensions to curvilinear coordinates and immersed boundary techniques for complex geometries \cite{chen2016vorticity,Poncet2009}. These solvers often rely on turbulence modeling strategies such as LES or RANS \cite{fox2012large,balachandar2010turbulent}, coupled with particle or interface tracking, which significantly increases computational complexity. In contrast, our formulation targets low-Reynolds-number viscous flows dominated by interfacial mechanics. By leveraging a vorticity–vector potential framework in time-dependent curvilinear coordinates \cite{chen2016vorticity}, the method reduces the governing equations to two Poisson problems, ensuring strict incompressibility without pressure computation and accommodating deformable boundaries without resorting to immersed boundary techniques \cite{Poncet2009}. This distinction highlights the complementary role of our scheme relative to existing solvers designed for turbulent multiphase flows.

Recent advances in the simulation of vesicle and cell dynamics in microfluidic flows include mesoscale frameworks \cite{noguchi2005shape} and particle-based approaches like dissipative particle dynamics \cite{fedosov2010multiscale}, as well as continuum phase-field models for red blood cell rheology\cite{LazaroSM1}. These state-of-the-art methods provide accurate physical descriptions but often involve complex implementations or high computational cost. In contrast, our approach introduces a simpler Poisson-based solver coupled with a phase-field description, enabling the incorporation of Helfrich bending and Cahn–Hilliard energies within a unified framework. This design prioritizes flexibility and ease of modification, making it a practical alternative for exploring membrane mechanics in confined flows. By openly sharing our code, we aim to complement existing high-performance methods and foster benchmarking efforts in this emerging area.

There is a clear gap in the literature regarding the use of vorticity-based formulations for deformable interfaces governed by Canham–Helfrich or Cahn–Hilliard energies in 3 dimensions. Most established approaches for vesicle and cell dynamics—such as those by Noguchi et al. (2005) \cite{noguchi2005shape}, Fedosov et al. (2010) \cite{fedosov2010multiscale}, and Lázaro et al. (2014) \cite{LazaroSM1}—rely on particle-based or continuum phase-field methods coupled with direct Navier–Stokes solvers, rather than vorticity-based schemes. To our knowledge, no existing work combines the advantages of vorticity–velocity formalism with phase-field models for membrane bending in 3 dimensions. Our method addresses this gap by introducing a simple Poisson-based solver that naturally integrates Helfrich bending forces and surface tension within a unified framework, offering a flexible alternative to more complex or computationally expensive techniques.

}

\section{Physical equations}
The following mathematical procedure could be taken for various interfacial free energies. 
For the derivation of the full model, we will be using a membrane free energy, the Canham-Helfrich energy. 
However, we will show results for a surface tension-free energy using the Cahn-Hilliard model, as it applies to different interfacial energies.

The membrane free energy will be written as,
\begin{equation}
F_{m}=\int_{\Gamma}\left(\frac{\kappa}{2}\left(2H-C_{0}\right)^{2}+\kappa_{G} K\right) dS  + \lambda_A A,
\label{eq:Fbcompl}
\end{equation}
{\color{black}for a membrane of bending and Gaussian rigidities $\kappa$ and $\kappa_G$, with mean and Gaussian curvatures $H$ and $K$}, spontaneous curvature $C_0$ and with an area Lagrange multiplier $\lambda_A$.
With this free energy, we can get the chemical potential as a functional derivative of the chosen free energy $\mu = \delta F / \delta \phi$.
This chemical potential is commonly used to write down the dynamic equation of the interface $\phi$ field
\begin{equation}
        \frac{\partial \phi }{\partial t} = M\nabla^2 \mu  - \vec v  \cdot  \nabla \phi \ , \label{dynamic}
\end{equation}
where $\vec v$ is the velocity field of the fluid, accounting for both the external and internal liquid from the cell/droplet, and  $M$ is a suitable mobility for the field $\phi$.

In addition, one needs to solve the actual fluid flow to be able to compute the evolution of the system.
We can compute the flow coupled to an immersed deformable object using the chemical potential, the vorticity $\vec \omega$, and the vector potential $\vec \psi$.
To account for this, we first define the vector potential related to the velocity field, $\vec v$, as
\begin{equation}
\vec v=\nabla\times\vec \psi,
    \label{eq:vecpot}
\end{equation}
and the vorticity as
\begin{equation}
    \vec \omega = \nabla \times \vec v.
    \label{eq:vort}
\end{equation}
Using these, we can write the following system of equations 
\begin{subequations} \label{eq:system}
\begin{empheq}{align}
    \partial_t \phi &= M( \nabla^2 \mu + \lambda_V) - (\nabla \times \vec{\psi})  \cdot  \nabla \phi \ , \label{gen2}
    \\
     \nabla^2 \vec{\omega} &=  \dfrac{1}{\eta} (\nabla \phi) \times (\nabla \mu), \label{gen1}
    \\
   \nabla^2 \vec{\psi} &= -\vec{\omega}, \label{gen3}
   \end{empheq}
\end{subequations}
where $\eta$ is the {\color{black}dynamic} viscosity of the fluid and $\lambda_V$ the volume Lagrange multiplier {\color{black}(details in Supplementary Materials)}. 
We arrived at this system of equations using the phase-field approach, which involves utilizing an order parameter called $\phi$.
{The no-slip condition at the deformable body, the Kinematic condition, the Stress balance, and the tangential stress are derived for the both Cahn-Hilliard and Canham-Helfrich equations in the Supplementary Materials.
These conditions are fulfilled by this flow-coupled phase field as can be seen by doing an asymptotic derivation of the effective boundary conditions. 
} 

{The Poisson equation for the vorticity (\ref{gen1}) comes from the curl of the Stokes equation
\begin{equation}
    0 = - \nabla P + \eta \nabla^2 \vec v - \phi \nabla \mu
\end{equation}
where the term $-\phi \nabla \mu$ accounts for the effect of the cell on the fluid.
This derivation can also be done using the Navier-Stokes equation if one is interested in the inertial contribution (see Supplementary Materials) by taking the curl of the incompressible Navier-Stokes equation. }
The Poisson equation (\ref{gen3}) is derived using curl properties, and vorticity can be expressed as
\begin{equation}
   \vec{\omega} =  \nabla \times \nabla \times \vec{\psi} \ \ \rightarrow  \ \  \vec{\omega} = \nabla (\nabla   \cdot   \vec{\psi}) - \nabla^2 \vec{\psi}.
    \label{eq:vort3d}
\end{equation}
Taking the gauge condition $\nabla \cdot \vec{\psi} = 0$~\cite{weinan1997finite}, a Poisson equation is derived from Eq. (\ref{eq:vort3d}), leading to a system of equations composed of two Poisson equations (\ref{gen1}) and (\ref{gen3}). 

 This methodology is a type of immersed boundary method that uses continuum fields to treat boundaries and surfaces as diffusive interfaces. To represent the selected free energy, we use the phase-field approach, either for bending with $C_0=0$~\cite{campelo06}
 \begin{equation}
    F_m[\phi] = \int_\Omega  \dfrac{3 \sqrt{2} \kappa}{8 \varepsilon^{3}}  (-\phi + \phi^3 -\varepsilon^2 \nabla^2 \phi)^2 dV + \lambda_A A[\phi].
\end{equation} 
{ \color{black}
To evaluate simulations that use this interfacial energy term, we will be using the bending characteristic time such as
\begin{equation}
    \tau_\kappa = \frac{\eta \, \mathcal{V}}{\kappa}, 
    \label{eq:tau_k}
\end{equation} 
where $\mathcal{V}$ is the volume of the cell.}

This formalism can be extended to other types of interfaces, like surface-tension-driven ones. 
We can model oil in water or other immiscible fluids by the use of the Cahn-Hilliard functional in place of the membrane one
\begin{equation}
    F_{CH}[\phi] = \int_\Omega  \Big( \frac{1}{4}(\phi^2 -1)^2 + \frac{\sigma}{2} | \nabla \phi|^2  \Big)  dV,
\end{equation} 
where $\sigma$ is the surface tension between the two fluids.
With this free energy, one can compute the new chemical potential $\mu = \delta F / \delta \phi$ and substitute that into equation (\ref{gen2}).
Therefore, changing from simulating one type of interface to another is easy to model and to implement numerically if you already have the code for another system.

\subsection{In plane shear}

We compute the in-plane stress at the interface, which we define as a tensor and intensity:
\begin{equation} \begin{array}{cc}
     \sigma_{ij}(x,y,z) = \eta ({\partial_j v_i + \partial_i v_j})  \\ \\
    \sigma_{t}(x,y,z) = \sqrt{{\sigma_{xy}}^2 + {\sigma_{yz}}^2 + {\sigma_{xz}}^2},
\end{array}
\end{equation}
where $\delta_i$ corresponds to the derivative with respect to the Cartesian coordinate $i$, which can be any from $(x,y,z)$.
This shear is especially useful for simulating cell membranes, as most cells have a skeleton or spring-like network attached to the membrane.
This skeleton can grant shear resistance to the membrane; thus, simulations in 2D cannot completely reflect the complexity of a cell surface.

\subsection{Boundary conditions} 
The system of equations \eqref{gen2}-\eqref{gen3} can accommodate any geometry through the boundary conditions. In this work, we take for all systems the no-slip boundary condition (BC) so that $\vec{v} = 0$ at the walls of the channel.
And thus, the BCs of the fluid are equivalent to the conditions without immersed bodies.
For example, for a cylindrical Poiseuille flow  $v_z^\text{P\,cy} = \Delta P/(4\eta L)\cdot(R^2-(r-R)^2)$ where $\Delta P$ represents the applied pressure to a cylindrical channel of length $L$ and radius $R$, we will introduce Dirichlet BCs in the channel walls that comply in the Cartesian space $(x,y,z)$ 
\begin{equation}
    \begin{split}
    \vec{\omega}\,^\text{P} & = \Big(  \dfrac{-\Delta P (y-R)}{2 \eta L }\  ,  \ \dfrac{\Delta P (x-R)}{2 \eta L} \  ,  \ 0\Big)  \\
    \vec{\psi}\,^\text{P} & = \dfrac{\Delta P}{4\eta L}\Big(  \big( \dfrac{y^3}{3} - \dfrac{R^2 y}{2} \big) \  ,  \ - \big( \dfrac{x^3}{3} - \dfrac{R^2 x}{2}\big)  \ ,  \ 0\Big) 
    \end{split}
    \label{eq:BCsPoiseuille}
\end{equation}
Obtaining the BC is easy, as one only has to compute $\vec \psi$ and $\vec \omega$ from the desired expression for $\vec v$. 
If they are not solvable analytically, one can always use a numerical solver to obtain the numerical value at the boundaries without bodies in the channel.
So, virtually any channel shape and flow type can be adapted to the method. 
As long as the channel shape is periodic, we will take {periodic boundary conditions} (PBC) in the flow direction to make the channel infinitely long.

The type of flow can be adapted by changing the boundary conditions. These BCs will basically cascade from the walls of the system to produce the desired kind of flow. 
We can use the analytical solution of other flows, like Couette, to simulate these microfluidic conditions.
The BCs for the Couette rectangular channel can be computed from the Couette flow velocity. 
For a moving wall in the $z$ direction, separated from a static wall in the direction $y$, where the height of the channel $h$ in the $y$ direction is so much smaller than one can take the width in the $x$ direction to be infinite
\begin{equation*}
    v_z(x,y) = v_{wall} \frac{y}{h}
\end{equation*}
will give the following BCs for vorticity and vector potential
\begin{equation}
    \begin{split}
    \vec{\omega}\,^\text{Co} & = v_{wall} \frac{y^2}{h},  \\
    \vec{\psi}\,^\text{Co} & = -  v_{wall} \frac{1}{h},
    \end{split}
    \label{eq:BCsCouete}
\end{equation}
equations that will be evaluated only at the walls, thus at $y=0$ and $y=h$.

\subsection{Numerical implementation}

The numerical implementation of our system of equations is done using finite differences and an unfitted interface methodology. 
Therefore, we define a fixed cubic grid that does not follow the interface or change over time.
In this grid, we will solve for the velocity fields and the field $\phi$. 
The exact position of the interface $\phi=0$ can be interpolated with good accuracy from the lattice results to plot the shape of the interface.

The {time evolution of the} field $\phi$ dynamic equation (\ref{gen2}) is solved using a forward Euler integration method {\color{black}using a step $\Delta t$. This method is sufficient for the low-Reynolds scope of this paper, where for a sufficiently small $\Delta t$ the Euler integration method will be stable}.
Then, {given that the flow is assumed non-inertial in the results of this paper, the} two Poisson equations (\ref{gen1}) and (\ref{gen3}) can be solved using an iterative method. 
The grid points are separated by the spatial resolution $\Delta x$, which we always take as equal in both dimensions $x$, $y$, and $z$.
The system is solved using a staggered scheme {\color{black}(also known as sequential segregated scheme or decoupled), where we solve each PDE successively, instead of simultaneously. This scheme} works properly for a small enough time increment $\Delta t$.
{Specifically, we perform these steps: (1) advance $\phi$ explicitly;  (2) solve the vorticity Poisson equation; (3) solve the stream-vector Poisson equation. All during the same time iteration and using  the same time step $\Delta t$. 
}

The implementation has been done in Python using an in-house code (freely available as an open source GitHub repository {\color{black}\cite{github3Dflow}}).

\section{Results and Discussion}

\begin{figure*}[h!]
\begin{center}
\includegraphics[width=.95\textwidth]{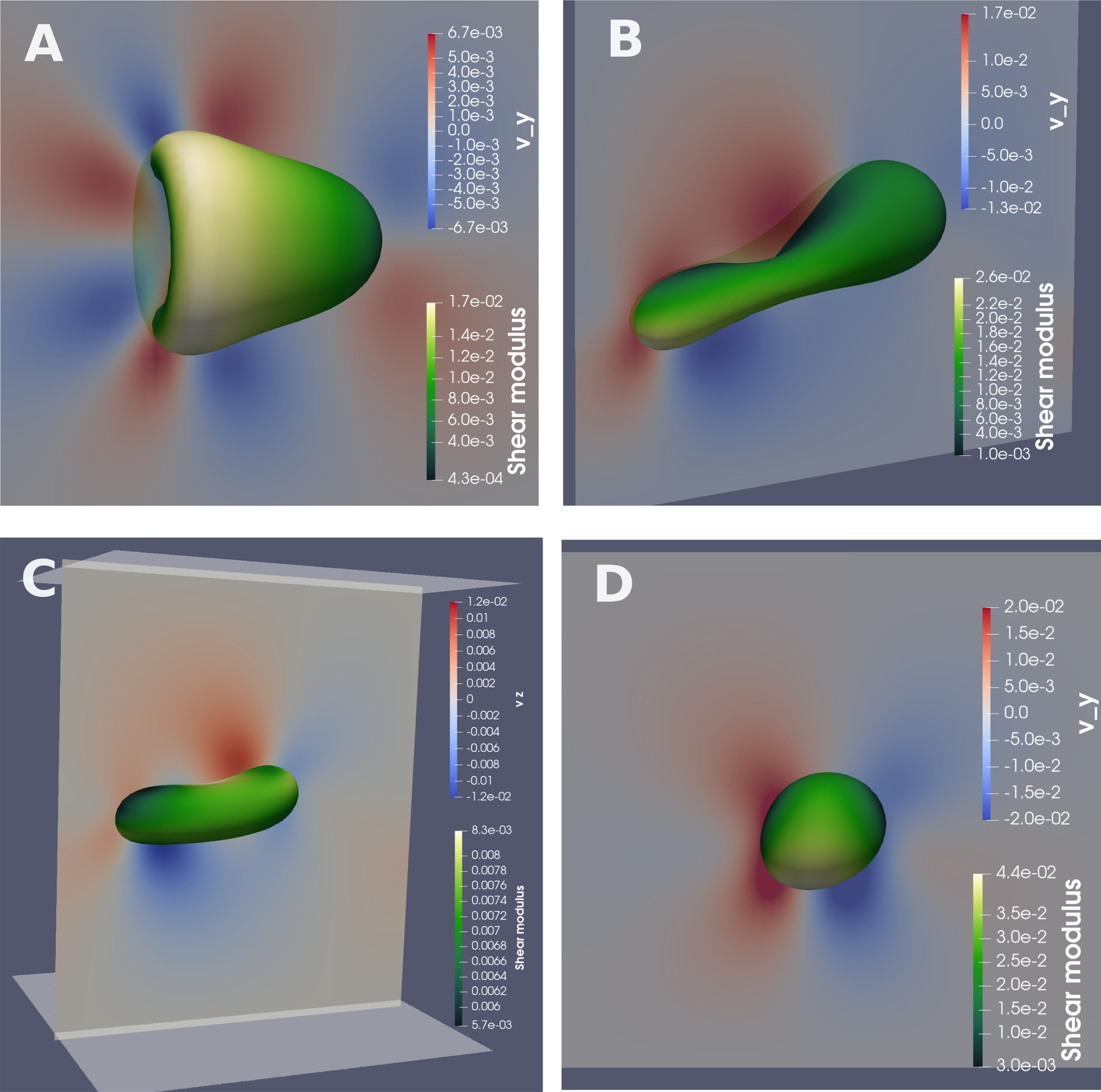}
\caption{First three plots correspond to a membrane inside a microfluidic flow. 
On the surfaces, we have plotted the shear. 
The presence of the obstacle also represented the deformation of the original fluid flow $y$ component.
\textbf{A,B} Membrane with Poiseuille both with $\kappa=1$ and a velocity in the center of the channel of $v=0.35\Delta x/ \Delta t$ but \textbf{A} starting centered in the channel and \textbf{B} outside the channel center. Results in   \textbf{C} show Helfrich membrane in a  Couette flow with $\kappa=1$ and a wall velocity of $v_{wall}=0.2$. \textbf{D} corresponds to an oil droplet in a Poiseuille flow starting outside of the center of the channel with a surface tension $\sigma=1$ and a central flow speed of $v=1.0$.}
\label{FIG:1}
\end{center}
\end{figure*}

A wide variety of combinations is possible for simulations; one can change the flow between Poiseuille and Couette or the interfacial free energy between a biological membrane and Cahn-Hilliard. 
On top of that, time-dependent flows like oscillatory flows can be simulated or change the viscosity only of either the outer fluid or the object's fluid, introducing a viscosity contrast.
In \textbf{FIG.}\ref{FIG:1}, a set of examples of what this methodology can be used for is represented.

This figure represents the main examples used in this paper, which will be biological membranes in Poiseuille and Couette flows, plus simulations of droplets using the Cahn-Hilliard energy functional as a workbench for the method.
Additional results and plots, especially on Cahn-Hilliard, can be found in the Supplementary Materials.

\subsection{Membranes in Poiseuille flows}

We start by discussing the results for the Canham-Helfrich membrane energy in Poiseuille flows. 
The reason is that there are many studies on the shape of RBCs in microfluidic flows.
Looking at (Fig.\ref{FIG:1} \textbf{A-B}), we can see that we recover known RBC shapes in low Reynolds flows, such as the parachute (Fig.\ref{FIG:1} \textbf{A}) or the slipper (Fig.\ref{FIG:1} \textbf{B}), well studied in the literature~\cite{shen2023anomalous}.

\begin{figure}[ht]
\begin{center}
\includegraphics[width=0.7\textwidth]{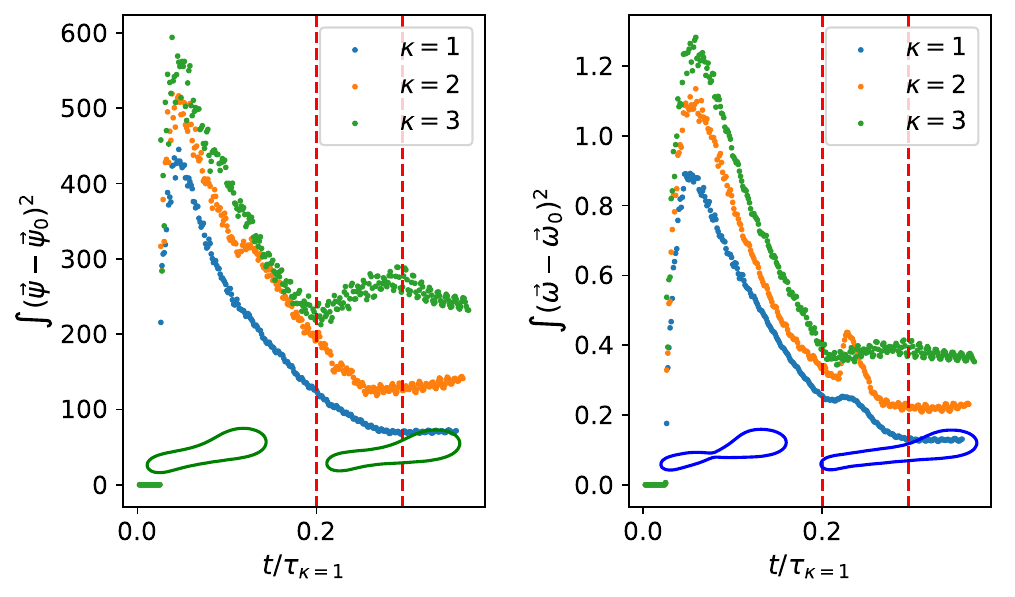} 
\caption{Observables computing the influence of the elastic object on the flow of fluid in the channel. 
Computed using the vorticity and stream vector, $\vec\omega$ and $\vec\xi$, from the simulations and subtracting $\vec \omega_0$ and $\vec\xi_0$, which are the vorticity and stream vector in the absence of the object.
These observables are computed at different bending rigidities $\kappa$, and the cross section of the cell is also represented for $\kappa=1$ at the left and $\kappa=3$ on the right, with the times of that cross section indicated in dotted lines.
}
\label{FIG:2}
\end{center}
\end{figure}

For the parachute, the maximum shear stress on the membrane is on the outer rim, while the front has a low shear due to the cylindrical symmetry. 
Meanwhile, a slipper-shaped vesicle has a much more interesting shear stress, concentrating most of the shear at the bottom back of the cell.

We compute a couple of observables we started studying in a previous work \cite{gallen2021red}. 
These observables are computed from the fluctuations of the vorticity and stream vector, $\vec\omega$ and $\vec\xi$, from the expected results in the absence of an object inside the fluid, $\vec \omega_0$ and $\vec\xi_0$,  a purely Poiseuille flow vorticity and stream vector.
So we take the integral of the squared fluctuations to study the influence of the object on the flow, as one can see in Figure \ref{FIG:2}.
We found these observables to be very reactive to small changes, and when an object stops deforming and moving around the channel, the values plateau as they act like a rigid body~\cite{gallen2021red}.

\subsection{ Effect of shape changes on our observables}
\begin{figure}[ht]
\begin{center}
\includegraphics[width=0.55\textwidth]{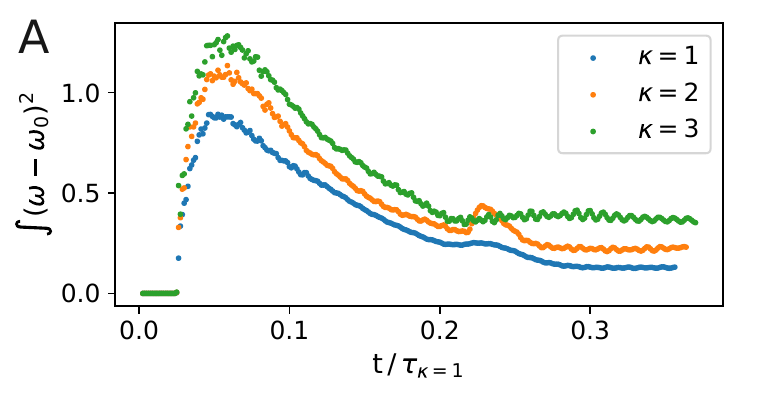} \\
\includegraphics[width=0.275\textwidth]{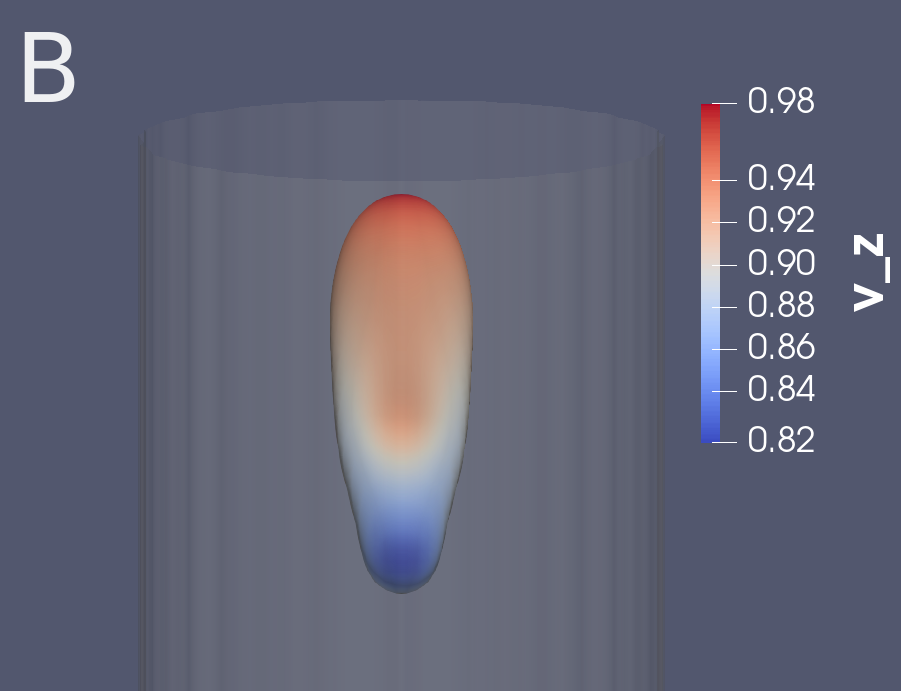} 
\includegraphics[width=0.25\textwidth]{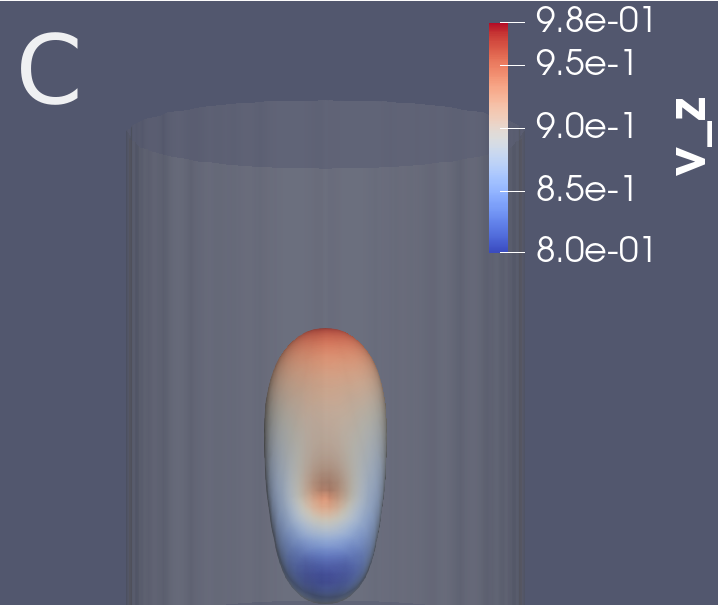    } 
\includegraphics[width=0.25\textwidth]{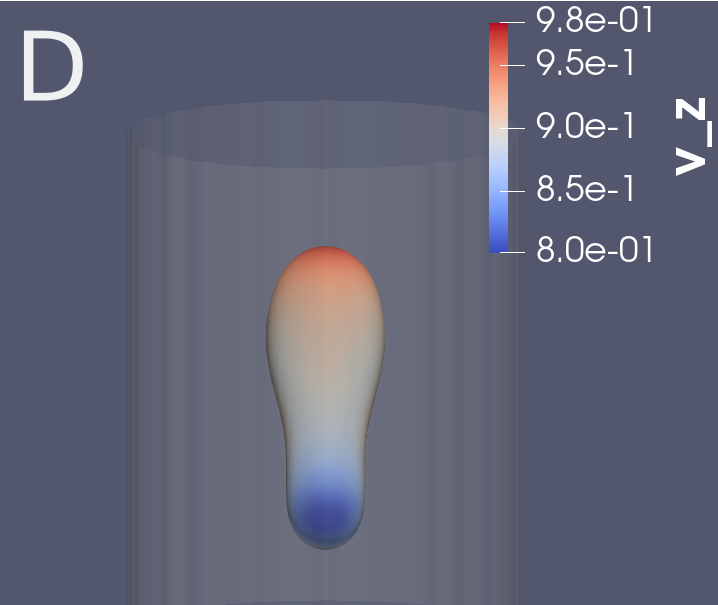    } 
\caption{(A) Shape changes around the spike in $\kappa=2$ around $t/\tau_{\kappa=1}=0.23$ show the disappearance of a dimple present before the peak. 
Bottom row 3D representations of the state around $t/\tau_{\kappa=1}=0.23$ for varying bending rigidities (B) $\kappa=1$; (C) $\kappa=2$; and  (D) $\kappa=3$.
For $\kappa=1$, the small dimple also appears, although less pronounced, while it does not appear for $\kappa=3$.}
\label{Fig:dimple}
\end{center}
\end{figure}

For example, inspection of the vorticity fluctuations for various bending modulus $\kappa$ from Figure \ref{FIG:2}.
For $\kappa=1$ and $\kappa=2$, we have two bumps during the relaxation related to the objects deforming to their steady state solutions. 
The bump is more pronounced for $\kappa=2$, and as we can see in Figure \ref{Fig:dimple} there is a disappearance of a dimple in the cell shape during the peak.
This happens both in $\kappa=1$ and $\kappa=2$.
So, {shape changes in the cell shape result in  changes in our observables.}
One could then use them as indicators for transformation and changes in the system, to know where to look.

{In Figure \ref{Fig:dimple} we can observe the lack of this dimple for $\kappa=3$. This makes physical sense as we expect that the dimples cannot appear in higher rigidities because it is a point of higher curvature, which a high bending rigidity will penalize strongly. The higher peak in our observable for $\kappa = 2$ rather than $\kappa = 1$ we expect it to mainly be simulation variance and noise. }

\subsection{Membranes in Couette flow and non-intertial lateral migration}

\begin{figure}[!htp]
\begin{center}
\includegraphics[width=0.9\textwidth]{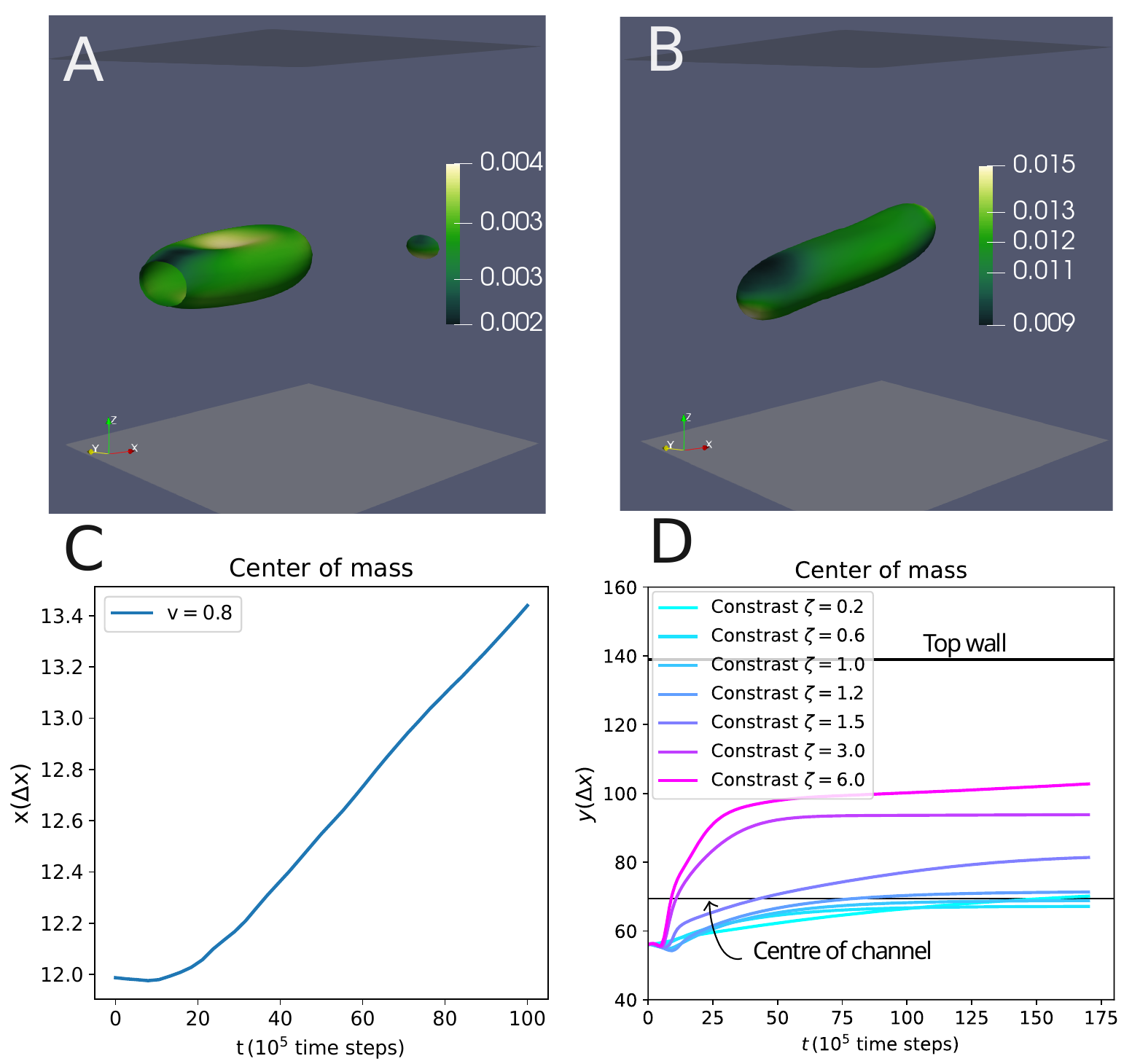}  
    \caption{
    (\textbf{A}) Final vesicle shape for $v_{wall}=0.2$, showing a relatively compact morphology. 
    (\textbf{B}) Final vesicle shape for $v_{wall}=0.8$, where the vesicle elongates significantly due to the higher shear. 
    Color bars in panels \textbf{A} and \textbf{B} correspond to the distribution of shear stress on the membrane, highlighting stronger gradients at higher wall velocity. 
    These results demonstrate that the temporal relaxation dynamics and steady-state configurations of vesicles in Couette flows depend sensitively on the imposed flow strength, with higher $v_{wall}$ producing longer transients and greater in-plane stress heterogeneity.
    \textit{  {Lateral migration of red blood cell (RBC)-shaped vesicles in Couette flow:} }
    (\textbf{C}) Three-dimensional simulation of a vesicle with diameter $40\Delta x$ in a channel of height $65\Delta x$, a viscosity contrast $\zeta=1$, and a $\Delta t = 10^{-4}$, corresponding to identical viscosities inside and outside the vesicle. The trajectory of the vesicle center of mass shows a clear drift toward the moving wall at $v_{wall}=0.8$. 
    (\textbf{D}) Two-dimensional simulations of vesicles with diameter $80\Delta x$ in a channel of height $140\Delta x$ at different viscosity contrasts $\zeta=\eta_{RBC}/\eta_{out}$ and a $\Delta t = 10^{-3}$. Each curve represents the migration trajectory of the vesicle center of mass, normalized by the channel height, starting from an initial offset from the channel centerline. Lower viscosity contrasts ($\zeta<1$) lead to slower or negligible migration, whereas higher contrasts ($\zeta \geq 1.5$) produce pronounced drift toward the top wall, where the vesicle stabilizes at distinct equilibrium positions. 
    }
\label{FIG:Couette}
\end{center}
\end{figure}

We simulated red blood cell (RBC)-shaped vesicles in a Couette flow, as shown in Figure \ref{FIG:1}\textbf{D}, to investigate {\color{black}the shape deformations as well as an effect specific to Couette flows called \textit{lateral migration}}. 
We see the influence of wall velocity on system dynamics where  higher wall speeds substantially delay the attainment of steady state shape as the final shape deformations are much greater for high speeds as seen in Figure \ref{FIG:Couette}\textbf{A},\textbf{B} where it depicts the corresponding vesicle morphologies at two representative shear rates. 
This delayed equilibration arises from the enhanced deformation imposed at larger shear, even though the intrinsic deformation timescale of the membrane remains nearly unchanged. Consequently, relaxation toward a stable configuration is prolonged under stronger flows. {More details can be seen in the Supplementary Materials.}

A known effect that occurs to objects immersed in a Couette flow is called \textit{lateral migration}, in which a lift force appears to act on the object perpendicular to the flow direction.
There are different sources, like the \textit{inertial lift} or Segré-Silberberg effect \cite{ho1974inertial}, a
\textit{non-inertial lift} force that seems to depend on the distance to the wall, the orientation and length of the particle, and the viscosity contrast \cite{olla1997lift,lorz2000weakly}.
Thus, the origins of this emerging lift force can be inertial or non-inertial; the latter would be the case here.

This effect is properly reproduced in both 2D (using the code from \cite{gallen2021red}) and 3D simulations of vesicles in Couette flows, represented in Figure \ref{FIG:Couette} \textbf{C, D}, demonstrating the capacity of the methodology and being a new tool to study this \textit{lateral migration} effect on cells in Couette flows.
{In Figure \ref{FIG:Couette} \textbf{C} we see the effect happening for a 3D simulation, while in Figure \ref{FIG:Couette} \textbf{D} we show how this effect changes in 2D with viscosity contrast $\zeta$, as 2D is much more convenient to study the parameter space that requires to run multiple simulations.
For a high enough viscosity contrast, the cell collides with the wall and it cannot keep moving upwards.}
The equilibrium position of the cell center of mass in the channel depends on parameters like viscosity contrast $\zeta$ or cell reduced volume (see supplementary materials).
The time taken for a cell to reach the steady state will then depend on the velocity of the flow and the distance between the starting position of the cell and the equilibrium steady state position.
An upper bound for this position will come from the cell reaching contact with the upper wall and having no more room to travel.

These results confirm that the driving velocity strongly controls the equilibrium position, the temporal pathway, and the membrane stress distribution. 
The stress maps in Figure \ref{FIG:Couette} panels \textbf{C} and \textbf{D} show how higher shear promotes larger in-plane tension gradients across the vesicle surface, which may in turn trigger morphological instabilities or transitions under more extreme conditions.
These results demonstrate the strong dependence of lateral migration on viscosity contrast and confirm that the framework captures both the qualitative trends and quantitative scaling of cell migration in Couette flows.
Together, these observations highlight the framework's robustness in capturing subtle dynamical features of vesicle migration in Couette flow. 




\section{Discussion and Conclusions}

The method we have introduced here offers several advantages for simulating three-dimensional incompressible flows interacting with deformable obstacles. Its simplicity in formulation and implementation makes it highly versatile and easy to adapt to diverse channel geometries and flow configurations, provided periodicity is maintained in the flow direction. Unlike lattice Boltzmann or other mesoscopic approaches that require proving consistency with continuum mechanics, our framework is directly based on classical fluid mechanics and avoids additional layers of justification.

{\color{black}
In three-dimensional incompressible flows with small Reynolds numbers, using vorticity and vector potential remains advantageous and the dynamics simplify considerably. At low Reynolds numbers, viscous forces dominate over inertial effects, which suppresses vortex stretching and tilting—the mechanisms that typically complicate 3D vorticity evolution. As a result, the vorticity equation reduces to a diffusion-dominated form, making the flow more predictable and easier to model. Expressing velocity as the curl of a divergence-free vector potential still enforces incompressibility automatically, while eliminating the pressure term from the governing equations. Although the full vector potential formulation requires handling gauge conditions, the absence of strong nonlinear interactions at low Reynolds numbers makes this approach particularly effective for analyzing deformable bodies in laminar regimes in three dimensions. 
}

{\color{black}In Pouiseuille flow we see that} by coupling the vorticity–stream-vector formulation with a phase-field description, the method flexibly incorporates different free-energy models for the interface. In particular, simulations using a bending-energy description reproduce experimentally observed red blood cell morphologies in microfluidic flows, such as discocyte and parachute shapes. This provides strong evidence of the framework’s physical fidelity. Moreover, the diagnostic observables we propose—integrals of squared vorticity and stream-vector fields—are remarkably sensitive to shape changes. They successfully detect transformations, continuous evolutions, and sudden morphological transitions, plateauing when the deformable body behaves as an effective rigid particle (see Fig.~3).

The Couette flow simulations further underscore the framework's robustness by capturing the lateral migration of vesicles. This phenomenon, for which a unified theoretical description remains elusive, is shown here to depend systematically on viscosity contrast, reduced volume, and initial placement. Our results demonstrate that the method can reliably quantify equilibrium offsets and migration dynamics, thus providing a practical computational tool for clarifying this long-standing problem in non-inertial flows. Extensions to inertial regimes would be a natural next step.

Importantly, the framework enables investigations that 2D models cannot, particularly in capturing in-plane shear dynamics essential for membrane mechanics. The phase-field approach allows straightforward integration of additional physical effects, including magnetic interactions in ferrofluids,  {on cells feeded with iron oxide nanoparticles \cite{mazuel2015magnetic}}, external body forces, and multi-vesicle suspensions, without significant computational overhead.

Finally, the method establishes a solid foundation for future developments. Incorporating viscoelastic suspending media, time-dependent flows such as oscillatory or transient regimes, and multi-body interactions are all well within reach. This adaptability ensures broad applicability to fluid–structure interaction problems across microfluidics, biological flows, and soft matter physics.

\subsection*{Supplementary Material}

Additional mathematical details and a derivation of the sharp-interface limit of the phase-field model are provided as Supplementary Material (PDF).

\subsection*{Acknowledgments}
\begin{acknowledgments}
A.H.M. acknowledges financial support from Ministerio de Ciencia e Innovación (MICINN, Spain) project PID2022-137994NB-I00 and "AGAUR" (Generalitat de Catalunya) under project 2021 SGR 00450. 
M.C. acknowledges financial support from Ministerio de Ciencia e Innovación (MICINN, Spain) project PID2022-140217NB-I00. 

\end{acknowledgments}

\bibliography{references}

@article{whitesides2006origins,
  title={The origins and the future of microfluidics},
  author={Whitesides, George M},
  journal={Nature},
  volume={442},
  number={7101},
  pages={368--373},
  year={2006},
  publisher={Nature Publishing Group UK London}
}

@article{baroud2010dynamics,
  title={Dynamics of microfluidic droplets},
  author={Baroud, Charles N and Gallaire, Francois and Dangla, R{\'e}mi},
  journal={Lab on a Chip},
  volume={10},
  number={16},
  pages={2032--2045},
  year={2010},
  publisher={Royal Society of Chemistry}
}

@article{tabeling2005introduction,
  title={Introduction to Microfluidics Oxford University Press},
  author={Tabeling, P},
  journal={) Book Introduction to Microfluidics Oxford University Press (Oxford, England ISBN, 2005, edn.)},
  pages={978--970},
  year={2005}
}

@article{sajeesh2014hydrodynamic,
  title={Hydrodynamic resistance and mobility of deformable objects in microfluidic channels},
  author={Sajeesh, P and Doble, M and Sen, AK},
  journal={Biomicrofluidics},
  volume={8},
  number={5},
  year={2014},
  pages = {054112},
  publisher={AIP Publishing}
}

@article{sajeesh2015microfluidic,
  title={A microfluidic device with focusing and spacing control for resistance-based sorting of droplets and cells},
  author={Sajeesh, P and Manasi, S and Doble, M and Sen, AK},
  journal={Lab on a Chip},
  volume={15},
  number={18},
  pages={3738--3748},
  year={2015},
  publisher={Royal Society of Chemistry}
}

@article{sajeesh2014particle,
  title={Particle separation and sorting in microfluidic devices: a review},
  author={Sajeesh, P and Sen, Ashis Kumar},
  journal={Microfluidics and nanofluidics},
  volume={17},
  number={1},
  pages={1--52},
  year={2014},
  publisher={Springer}
}

@article{shen2024merged,
  title={Merged and alternating droplets generation in double T-junction microchannels using symmetrically inserted capillaries},
  author={Shen, Feng and Zhang, Yuedong and Li, Chunyou and Pang, Yan and Liu, Zhaomiao},
  journal={Microfluidics and Nanofluidics},
  volume={28},
  number={5},
  pages={29},
  year={2024},
  publisher={Springer}
}

@article{shen2023anomalous,
  title={Anomalous diffusion of deformable particles in a honeycomb network},
  author={Shen, Zaiyi and Plourabou{\'e}, Franck and Lintuvuori, Juho S and Zhang, Hengdi and Abbasi, Mehdi and Misbah, Chaouqi},
  journal={Physical Review Letters},
  volume={130},
  number={1},
  pages={014001},
  year={2023},
  publisher={APS}
}

@article{mauer2018,
  title={Flow-induced transitions of red blood cell shapes under shear},
  author={Mauer, Johannes and Mendez, Simon and Lanotte, Luca and Nicoud, Franck and Abkarian, Manouk and Gompper, Gerhard and Fedosov, Dmitry A},
  journal={Physical Review Letters},
  volume={121},
  number={11},
  pages={118103},
  year={2018},
  publisher={APS}
}

@article{ho1974inertial,
  title={Inertial migration of rigid spheres in two-dimensional unidirectional flows},
  author={Ho, BP and Leal, LG0284},
  journal={Journal of fluid mechanics},
  volume={65},
  number={2},
  pages={365--400},
  year={1974},
  publisher={Cambridge University Press}
}

@article{guckenberger2017theory,
  title={Theory and algorithms to compute Helfrich bending forces: a review},
  author={Guckenberger, Achim and Gekle, Stephan},
  journal={Journal of Physics: Condensed Matter},
  volume={29},
  number={20},
  pages={203001},
  year={2017},
  publisher={IOP Publishing}
}

@article{gallen2021red,
  title={Red blood cells in low Reynolds number flow: A vorticity-based characterization of shapes in two dimensions},
  author={Gallen, Andreu F and Castro, Mario and Hernandez-Machado, Aurora},
  journal={Soft Matter},
  volume={17},
  number={42},
  pages={9587--9594},
  year={2021},
  publisher={Royal Society of Chemistry}
}

@article{Poncet2009,
  title={Analysis of an immersed boundary method for three-dimensional flows in vorticity formulation},
  author={Poncet, Philippe},
  journal={Journal of Computational Physics},
  volume={228},
  number={19},
  pages={7268--7288},
  year={2009},
  publisher={Elsevier}
}

@article{Olshanskii2010,
  title={Velocity--vorticity--helicity formulation and a solver for the Navier--Stokes equations},
  author={Olshanskii, Maxim A and Rebholz, Leo G},
  journal={Journal of Computational Physics},
  volume={229},
  number={11},
  pages={4291--4303},
  year={2010},
  publisher={Elsevier}
}

@article{meitz2000compact,
  title={A compact-difference scheme for the Navier--Stokes equations in vorticity--velocity formulation},
  author={Meitz, Hubert L and Fasel, Hermann F},
  journal={Journal of Computational Physics},
  volume={157},
  number={1},
  pages={371--403},
  year={2000},
  publisher={Elsevier}
}

@article{fox2012large,
  title={Large-eddy-simulation tools for multiphase flows},
  author={Fox, Rodney O},
  journal={Annual Review of Fluid Mechanics},
  volume={44},
  number={1},
  pages={47--76},
  year={2012},
  publisher={Annual Reviews}
}

@article{chen2016vorticity,
  title={Vorticity vector-potential method for 3D viscous incompressible flows in time-dependent curvilinear coordinates},
  author={Chen, Yu and Xie, Xilin},
  journal={Journal of Computational Physics},
  volume={312},
  pages={50--81},
  year={2016},
  publisher={Elsevier}
}

@article{balachandar2010turbulent,
  title={Turbulent dispersed multiphase flow},
  author={Balachandar, S and Eaton, John K},
  journal={Annual review of fluid mechanics},
  volume={42},
  number={1},
  pages={111--133},
  year={2010},
  publisher={Annual Reviews}
}

@article{weinan1997finite,
  title={Finite difference methods for 3D viscous incompressible flows in the vorticity--vector potential formulation on nonstaggered grids},
  author={Weinan, E and Liu, Jian-Guo},
  journal={Journal of Computational Physics},
  volume={138},
  number={1},
  pages={57--82},
  year={1997},
  publisher={Elsevier}
}

@article{mazuel2015magnetic,
  title={Magnetic flattening of stem-cell spheroids indicates a size-dependent elastocapillary transition},
  author={Mazuel, Francois and Reffay, Myriam and Du, Vicard and Bacri, Jean-Claude and Rieu, Jean-Paul and Wilhelm, Claire},
  journal={Physical review letters},
  volume={114},
  number={9},
  pages={098105},
  year={2015},
  publisher={APS}
}

@article{lanotte2016,
  title={Red cells’ dynamic morphologies govern blood shear thinning under microcirculatory flow conditions},
  author={Lanotte, Luca and Mauer, Johannes and Mendez, Simon and Fedosov, Dmitry A and Fromental, Jean-Marc and Claveria, Viviana and Nicoud, Franck and Gompper, Gerhard and Abkarian, Manouk},
  journal={Proceedings of the National Academy of Sciences},
  volume={113},
  number={47},
  pages={13289--13294},
  year={2016},
  publisher={National Acad Sciences}
}

@misc{github3Dflow,
  author = {A F Gallen},
  title = {MemPhaseFlow3D},
  year = {2025},
  publisher = {GitHub},
  journal = {GitHub repository},
  howpublished = {\url{https://github.com/fdzgallen/MemPhaseFlow3D/}}  
}

@article{campelo06,
  title={Dynamic model and stationary shapes of fluid vesicles},
  author={Campelo, Felix and Hernandez-Machado, Aurora},
  journal={The European Physical Journal E},
  volume={20},
  number={1},
  pages={37--45},
  year={2006},
  publisher={Springer}
}

@article{kaoui2011,
  title={Complexity of vesicle microcirculation},
  author={Kaoui, B and Tahiri, N and Biben, T and Ez-Zahraouy, H and Benyoussef, A and Biros, G and Misbah, C},
  journal={Physical Review E},
  volume={84},
  number={4},
  pages={041906},
  year={2011},
  publisher={APS}
}

@article{Lazaro2019,
  title={Collective behavior of red blood cells in confined channels},
  author={L{\'a}zaro, Guillermo R and Hern{\'a}ndez-Machado, Aurora and Pagonabarraga, Ignacio},
  journal={The European Physical Journal E},
  volume={42},
  number={4},
  pages={1--9},
  year={2019},
  publisher={Springer Berlin Heidelberg}
}

@article{noguchi2005shape,
  title={Shape transitions of fluid vesicles and red blood cells in capillary flows},
  author={Noguchi, Hiroshi and Gompper, Gerhard},
  journal={Proceedings of the National Academy of Sciences},
  volume={102},
  number={40},
  pages={14159--14164},
  year={2005},
  publisher={National Academy of Sciences}
}

@article{LazaroSM1,
	title={Rheology of red blood cells under flow in highly confined microchannels: I. effect of elasticity},
	author={L{\'a}zaro, Guillermo R and Hern{\'a}ndez-Machado, Aurora and Pagonabarraga, Ignacio},
	journal={Soft Matter},
	volume={10},
	number={37},
	pages={7195--7206},
	year={2014},
	publisher={Royal Society of Chemistry}
}

@article{lorz2000weakly,
  title={Weakly adhering vesicles in shear flow: Tanktreading and anomalous lift force},
  author={Lorz, B and Simson, R and Nardi, J and Sackmann, E},
  journal={Europhysics Letters},
  volume={51},
  number={4},
  pages={468},
  year={2000},
  publisher={IOP Publishing}
}

@article{olla1997lift,
  title={The lift on a tank-treading ellipsoidal cell in a shear flow},
  author={Olla, Piero},
  journal={Journal de Physique II},
  volume={7},
  number={10},
  pages={1533--1540},
  year={1997},
  publisher={EDP Sciences}
}

@article{fedosov2010multiscale,
  title={A multiscale red blood cell model with accurate mechanics, rheology, and dynamics},
  author={Fedosov, Dmitry A and Caswell, Bruce and Karniadakis, George Em},
  journal={Biophysical journal},
  volume={98},
  number={10},
  pages={2215--2225},
  year={2010},
  publisher={Elsevier}
}

@article{aouane2014vesicle,
  title={Vesicle dynamics in a confined Poiseuille flow: from steady state to chaos},
  author={Aouane, Othmane and Thi{\'e}baud, Marine and Benyoussef, Abdelilah and Wagner, Christian and Misbah, Chaouqi},
  journal={Physical Review E},
  volume={90},
  number={3},
  pages={033011},
  year={2014},
  publisher={APS}
}

@article{eggleton1998large,
  title={Large deformation of red blood cell ghosts in a simple shear flow},
  author={Eggleton, Charles D and Popel, Aleksander S},
  journal={Physics of fluids},
  volume={10},
  number={8},
  pages={1834--1845},
  year={1998},
  publisher={American Institute of Physics}
}

@article{aidun2010lattice,
  title={Lattice-Boltzmann method for complex flows},
  author={Aidun, Cyrus K and Clausen, Jonathan R},
  journal={Annual review of fluid mechanics},
  volume={42},
  pages={439--472},
  year={2010},
  publisher={Annual Reviews}
}

@article{Zhao2011TheDO,
  title={The dynamics of a vesicle in a wall-bound shear flow},
  author={Hong Zhao and Andrew P. Spann and Eric S. G. Shaqfeh},
  journal={Physics of Fluids},
  year={2011},
  volume={23},
  pages={121901},
  url={https://api.semanticscholar.org/CorpusID:120085753}
}

@article{Rallabandi,
author = {Rallabandi, Bhargav},
title = {Fluid-Elastic Interactions Near Contact at Low Reynolds Number},
journal = {Annual Review of Fluid Mechanics},
volume = {56},
number = {1},
pages = {491-519},
year = {2024},
doi = {10.1146/annurev-fluid-120720-024426},

URL = { 
    
        https://doi.org/10.1146/annurev-fluid-120720-024426
    
    

},
eprint = { 
    
        https://doi.org/10.1146/annurev-fluid-120720-024426
    
    

}
}
\bibliographystyle{rsc}
\clearpage


\section*{Supplementary material}

# Supplementary Materials: Deformable bodies in a 3-dimensional viscous flow: Vorticity-Stream vector formulation

Andreu F. Gallen[1],* Joan Muñoz Biosca[2], Mario Castro[3], and Aurora Hernandez-Machado[2,4]
[1]*Center for Interdisciplinary Research in Biology (CIRB),
Collège de France, CNRS, Inserm, Université PSL, Paris, France,*
[2]*Facultat de Física, Universitat de Barcelona, Diagonal 645, 08028 Barcelona, Spain,*
[3]*Instituto de Investigación Tecnológica (IIT) and Grupo Interdisciplinar de Sistemas Complejos (GISC),
Universidad Pontificia Comillas, Madrid, E28015, Spain, and*
[4]*Institute of Nanoscience and Nanotechnology (IN2UB), 08028 Barcelona, Spain.* *

## MATCHED ASYMPTOTIC DERIVATION OF EFFECTIVE BOUNDARY CONDITIONS

We consider a phase-field model for a membrane described by the Cahn-Hilliard free energy in the Stokes limit. The system in the diffuse interface region (inner region) is given by:

$$\partial_t \phi = M \nabla^2 \mu - \vec{v} \cdot \nabla \phi, \tag{1a}$$

$$0 = -\nabla p + \eta \nabla^2 \vec{v} - \phi \nabla \mu, \tag{1b}$$

$$\mu = \phi^3 - \phi - \varepsilon^2 \nabla^2 \phi, \tag{1c}$$

where $\phi$ is the phase field, $\mu$ the chemical potential, $\vec{v}$ the velocity, $p$ the pressure, $\eta$ dynamic viscosity, $M$ the mobility, and $\varepsilon$ the interface width parameter, related (if $\phi$ is transformed into a physical density) with the surface tension, $\sigma$.

### Coordinate system and scaling

Let the sharp interface be located at $r = 0$, with $r > 0$ representing the exterior fluid (e.g., aqueous phase) and $r < 0$ the interior (e.g., lipid phase). Introduce a stretched coordinate across the diffuse interface:

$$z = \frac{r}{\varepsilon}, \tag{2}$$

where $\varepsilon$ is the small parameter representing the interface width. The normal vector $\vec{n}$ points from the interior to the exterior. The gradient and Laplacian operators become:

$$\nabla = \vec{n} \partial_r + \nabla_s = \frac{1}{\varepsilon} \vec{n} \partial_z + \nabla_s, \tag{3}$$

$$\nabla^2 = \frac{1}{\varepsilon^2} \partial_{zz} + -\frac{2H}{\varepsilon} \partial_z + \nabla_s^2 + \mathcal{O}(\varepsilon), \tag{4}$$

where $\nabla_s$ denotes the surface gradient, and $H = -\frac{1}{2}\nabla \cdot \vec{n}$ is the mean curvature (half the sum of principal curvatures, with sign convention such that $H > 0$ for a sphere with outward normal).

We assume the following asymptotic expansions in the inner region:

$$\phi = \phi_0(z) + \varepsilon \phi_1(z,s) + \varepsilon^2 \phi_2(z,s) + \cdots, \tag{5}$$

$$\mu = \mu_0(z) + \varepsilon \mu_1(z,s) + \varepsilon^2 \mu_2(z,s) + \cdots, \tag{6}$$

$$\vec{v} = \vec{v}_0(z,s) + \varepsilon \vec{v}_1(z,s) + \cdots, \tag{7}$$

$$p = \frac{1}{\varepsilon} p_{-1}(z,s) + p_0(z,s) + \varepsilon p_1(z,s) + \cdots. \tag{8}$$

The pressure scaling $p_{-1}$ is introduced to balance the leading-order terms from the chemical potential gradient.

### Leading order analysis

The leading-order chemical potential satisfies:

$$-\sigma\partial_{zz}\phi_0 + \phi_0^3 - \phi_0 = 0 \tag{9}$$

The equilibrium profile satisfying $\phi_0(\pm\infty) = \pm 1$ is:

$$\phi_0(z) = \tanh\left(\frac{z}{\sqrt{2\sigma}}\right). \tag{10}$$

At order $\mathcal{O}(\varepsilon^{-2})$ from the normal component of (1b):

$$-\partial_z p_{-1} + \eta\partial_{zz} v_{0n} - \phi_0\partial_z \mu_{-1} = 0. \tag{11}$$

Since $\mu_0 = 0$ at equilibrium, $\partial_z \mu_0 = 0$, and matching to bounded outer pressure requires $p_{-1}$ to be constant. Thus, the viscous term $\eta\partial_{zz} v_{0n}$ must also vanish at this order, consistent with $\vec{v}_0$ being independent of $z$ in leading order.

From (1c) at next order

$$\mu_0 = (3\phi_0^2 - 1)\phi_1 - \sigma(\partial_{zz}\phi_1 + (-2H)\,\partial_z\phi_0). \tag{12}$$

Multiplying by $\partial_z\phi_0$ and integrating over $z$ from $-\infty$ to $\infty$, using that $\int_{-\infty}^{\infty}\partial_z\phi_0\,dz = 2$, we obtain:

$$\mu_0 \int_{-\infty}^{\infty} \partial_z\phi_0\,dz = +2\sigma H \int_{-\infty}^{\infty} (\partial_z\phi_0)^2 dz, \tag{13}$$

so

$$\mu_0 = \sigma H \int_{-\infty}^{\infty} (\partial_z\phi_0)^2 dz. \tag{14}$$

Evaluating the integral $\int_{-\infty}^{\infty}(\partial_z\phi_0)^2 dz = \frac{2\sqrt{2}}{3\sqrt{\sigma}}$ gives:

$$\mu_0 = \frac{2\sqrt{2\sigma}}{3}H = \gamma H, \tag{15}$$

where $\gamma = \frac{2\sqrt{2\sigma}}{3}$ is the surface tension. This is the Gibbs-Thomson relation: the leading-order chemical potential is proportional to the mean curvature.

### First-order analysis and stress condition

At $\mathcal{O}(1)$ from the Stokes equation (1b):

$$-\nabla_s p_{-1} - \vec{n}\partial_z p_0 + \eta(\partial_{zz}\vec{v}_0 + \partial_z\vec{v}_0) - \phi_0\nabla_s\mu_0 - \phi_0\vec{n}\partial_z\mu_1 - \phi_1\vec{n}\partial_z\mu_0 = 0. \tag{16}$$

Since $\mu_0$ is constant along the interface, $\nabla_s\mu_0 = 0$. The normal component at $\mathcal{O}(1)$ gives:

$$-\partial_z p_0 + \eta(\partial_{zz} v_{0n} + (-2H)\,\partial_z v_{0n}) - \phi_0\partial_z\mu_1 = 0. \tag{17}$$

Integrating across the interface and using matching conditions yields the jump in normal stress.

The leading-order tangential Stokes equation is at $\mathcal{O}(\varepsilon^{-2})$:

$$\eta\partial_{zz}\vec{v}_{0t} = 0, \tag{18}$$

so $\vec{v}_{0t} = \vec{A}(s)z + \vec{B}(s)$. Matching to bounded outer solutions as $z \to \pm\infty$ forces $\vec{A}(s) = 0$, hence $\vec{v}_{0t} = \vec{B}(s)$ is constant across the interface, and matching implies

$$\vec{v}_t^+ = \vec{v}_t^-,$$

so the tangential velocity is continuous.

The shear stress jump arises at $\mathcal{O}(1)$ and involves $\nabla_s p_0$ and Marangoni terms if $\mu_0$ varies along the interface. For constant $\mu_0$, the tangential stress is continuous.

The normal stress balance is obtained by matching the $\mathcal{O}(1)$ inner pressure to the outer solutions. The jump in outer pressure is given by

$$[p] = 2\eta[\partial_n v_n] + 2\gamma H, \tag{19}$$

where the surface tension is $\gamma = \frac{2\sqrt{2\sigma}}{3}$, consistent with the standard Laplace law (pressure higher on the concave side).

**Deformable bodies in a 3-dimensional viscous flow: Vorticity-Stream vector formulation**

### Kinematic condition

From the phase-field equation (1a) at $\mathcal{O}(\varepsilon^{-1})$:

$$0 = M\partial_{zz}\mu_0 - v_{0n}\partial_z\phi_0. \tag{20}$$

Since $\mu_0 = 0$ is constant, $\partial_{zz}\mu_{-1} = 0$, so $v_{0n}\partial_z\phi_0 = 0$, implying $v_{0n}$ is constant across the interface. Matching to outer solutions gives $v_{0n} = \vec{v}^{\pm} \cdot \vec{n}$, so the normal velocity is continuous.

At $\mathcal{O}(1)$, using $\partial_t\phi_0 = -V_n\partial_z\phi_0$ and integrating against $\partial_z\phi_0$, one recovers the kinematic condition:

$$V_n = \vec{v} \cdot \vec{n}. \tag{21}$$

### Effective boundary conditions on the sharp interface

Collecting the results, the matched asymptotic analysis yields the following effective boundary conditions on the sharp interface $\Gamma$:

1. **Continuity of velocity:**

$$[\vec{v}] = 0. \tag{22}$$

2. **Kinematic condition:**

$$V_n = \vec{v} \cdot \vec{n}. \tag{23}$$

3. **Stress balance:**

$$[\boldsymbol{\sigma}] \cdot \vec{n} = 2\gamma H \, \vec{n} - \nabla_s \gamma, \tag{24}$$

where $\boldsymbol{\sigma} = -p\mathbf{I} + \eta(\nabla\vec{v} + (\nabla\vec{v})^{\mathrm{T}})$ is the Newtonian stress tensor, and $\gamma = \frac{2\sqrt{2\sigma}}{3}$ is the surface tension. For constant $\gamma$, this reduces to the Laplace pressure condition:

$$[p] = 2\eta[\partial_n v_n] + 2\gamma H. \tag{25}$$

4. **Tangential stress:** For a clean interface (no Marangoni effects),

$$[\sigma_{nt}] = 0, \tag{26}$$

where $\sigma_{nt}$ denotes the tangential component of the stress.

These conditions, together with the Stokes equations in the bulk phases, form the sharp-interface model corresponding to the diffuse-interface Stokes–phase-field system.

### THE HELFRICH CASE

For the Helfrich bending energy in the phase-field formulation:

$$F_m[\phi] = \int_\Omega \frac{3\sqrt{2}\kappa}{8\varepsilon^3}(-\phi + \phi^3 - \varepsilon^2\nabla^2\phi)^2 dV,$$

where $\kappa$ is the bending modulus, the chemical potential contains higher-order derivatives:

$$\mu = \frac{3\sqrt{2}\kappa}{4\varepsilon^3}(-\phi + \phi^3 - \varepsilon^2\nabla^2\phi)(-1 + 3\phi^2 - \varepsilon^2\nabla^2).$$

Performing matched asymptotics with this $\mu$, the normal stress jump condition gains additional bending terms:

$$[p] = 2\eta[\partial_n v_n] + 2\gamma H + \kappa\left(2\Delta_s H + 4H(H^2 - K)\right) - 2\lambda_A H, \tag{27}$$

where $\Delta_s$ is the surface Laplacian, $H$ the mean curvature, $K$ the Gaussian curvature, and $\lambda_A$ the area Lagrange multiplier. The bending modulus $\kappa$ appears in the higher-order curvature terms, representing the membrane's resistance to bending.

The velocity boundary conditions are universal for clean fluid-fluid interfaces because:

**Deformable bodies in a 3-dimensional viscous flow: Vorticity-Stream vector formulation**

- **No-slip**: The phase-field $\phi$ is advected by the fluid velocity $\vec{v}$ via $\partial_t \phi + \vec{v} \cdot \nabla \phi = \cdots$. In the sharp-interface limit, this implies the interface moves with the fluid velocity, requiring continuity of $\vec{v}$ across the interface.
- **Kinematic condition**: This is equivalent to mass conservation for an interface without phase change. It emerges from the phase-field equation at $O(1)$ after integration across the interface.
- These conditions hold regardless of the specific free energy because they stem from the advection term $-\vec{v} \cdot \nabla \phi$ in the phase-field equation, which is common to all models.

The stress conditions, however, depend on the free energy because:

- The stress jump $[\boldsymbol{\sigma}] \cdot \vec{n}$ equals the functional derivative of the free energy with respect to interface deformation.
- For Cahn-Hilliard: $\delta F_{CH}/\delta \Gamma = 2\gamma H \, \vec{n}$ (only tension)
- For Helfrich: $\delta F_m/\delta \Gamma = \left[2\gamma H + \kappa(2\Delta_s H + 4H(H^2 - K)) - 2\lambda_A H\right] \vec{n}$ (tension + bending)
- Thus, the bending modulus $\kappa$ appears in the normal stress balance but not in the velocity conditions.

**PHYSICAL EQUATIONS**

We start with an interfacial free energy, for example a membrane free energy

$$F_m = \int_\Gamma \left(\frac{\kappa}{2}(2H - C_0)^2 + \kappa_G K\right) dS + \lambda_A A, \tag{28}$$

with mean curvature $H$, spontaneous curvature $C_0$ and with an area Lagrange multiplier $\lambda_A$. With this free energy we can get the chemical potential as a functional derivative of the chosen free energy $\mu = \delta F/\delta \phi$. Using the chemical potential we can compute the flow coupled to an immersed deformable object using the following system of equations

$$\partial_t \phi = M(\nabla^2 \mu + \lambda_V) - (\nabla \times \vec{\psi}) \cdot \nabla \phi, \tag{29a}$$

$$\nabla^2 \vec{\omega} = \frac{1}{\eta}(\nabla \phi) \times (\nabla \mu), \tag{29b}$$

$$\nabla^2 \vec{\psi} = -\vec{\omega}, \tag{29c}$$

where $\vec{\omega}$ is the vorticity, $\eta$ the viscosity of the fluid, $\mu$ the chemical potential, $\vec{\psi}$ the vector potential, $M$ a suitable mobility and $\lambda_V$ the volume Lagrange multiplier.

To reach this system of equations, we are using the phase field approach using an order parameter $\phi$. The phase field methodology is a type of immersed boundary method written using continuum fields, where boundaries and surfaces are modelled as diffusive interfaces. We take the phase field representation of the chosen free energy, whether bending with $C_0 = 0$ [1]

$$F_m[\phi] = \int_\Omega \frac{3\sqrt{2}\kappa}{8\varepsilon^3}(-\phi + \phi^3 - \varepsilon^2 \nabla^2 \phi)^2 dV + \lambda_A A[\phi], \tag{30}$$

or a surface tension energy using the Cahn-Hilliard free energy

$$F_{CH}[\phi] = \int_\Omega \left(\frac{1}{4}(\phi^2 - 1)^2 + \frac{\sigma}{2}|\nabla \phi|^2\right) dV. \tag{31}$$

The Lagrange multiplier can be computed via the penalty method, where the local surface area functional $A[\phi]$ [2, 3]

$$A[\phi] = \frac{3}{2\sqrt{2}} \int_\Omega \left(\frac{\epsilon}{2}|\nabla \phi|^2 + \frac{1}{4\epsilon}\left(\phi^2 - 1\right)^2\right) dV. \tag{32}$$

The dynamic equation for $\phi$ before the introduction of the hydrodynamic contribution is

$$\frac{\partial \phi}{\partial t} = M(\nabla^2 \mu + \lambda_V \frac{\delta V[\phi]}{\delta \phi}) = M(\nabla^2 \frac{\delta F}{\delta \phi} + \lambda_V), \tag{33}$$

**Deformable bodies in a 3-dimensional viscous flow: Vorticity-Stream vector formulation**

where $\delta/\delta\phi$ is the functional derivative defined in [4] as

$$\mu = \frac{\delta F[\phi]}{\delta\phi} = \frac{\partial F}{\partial\phi} - \nabla \cdot \frac{\partial F}{\partial(\nabla\phi)} + \nabla^2 \frac{\partial F}{\partial(\nabla^2\phi)}, \quad (34)$$

which when computed for the membrane energy gives us

$$\frac{\delta F_M[\phi]}{\delta\phi} = \frac{3\sqrt{2}\kappa}{8\epsilon^3}\Big((3\phi^2 - 1)(-\phi + \phi^3 - \epsilon^2\nabla^2\phi) - \epsilon^2\nabla^2(-\phi + \phi^3 - \epsilon^2\nabla^2\phi)\Big) - \gamma_A \nabla^2\phi. \quad (35)$$

The Lagrange multiplier $\lambda_V$ is not strictly necessary as we write this dynamic equation as a diffusive equation. It is mainly add for very high speed flows so that the conservation is ensured even at more extreme flows. This volume Lagrange multiplier $\lambda_V$ is adopted from the same penalty method.

### Time-dependency and inertial flow.

Using this formulation one can easily adapt the fluid flow to any flow profile that has an analytical expression. For example, one can use a flow velocity profile that is time-dependent. Like a flow that gets faster or slower over time or an oscillating flow. The system of equations (29) needs no change, the only thing one has to change for this are the boundary conditions.

Another possibility is to use the Navier-Stokes equation instead of its non-inertial counterpart. This leads us to the following system of equations

$$\begin{aligned} \partial_t \phi &= M(\nabla^2 \mu + \lambda_V) - (\nabla \times \vec{\psi}) \cdot \nabla\phi\,, \\ \frac{\rho}{\eta}\partial_t \vec{\omega} - \nabla^2 \vec{\omega} &= -\frac{1}{\eta}(\nabla\phi) \times (\nabla\mu), \\ \nabla^2 \vec{\psi} &= -\vec{\omega}. \end{aligned} \quad (36)$$

The only change we find in the final system of equations is in the vorticity equation, which instead of being a Poisson equation, one reaches the thermal conduction equation. However, we will focus this paper on the implementation for non-inertial low-Reynolds flows.

### Lagrange multipliers

Finally, the enclosed volume, which is related to the pressure gradient between the internal and external environment of the cell, has to be maintained. Taking the Model-B-like dynamics, the dynamic Eq. (29a) ensures conservation of the volume, which is defined by [2] as

$$V[\phi] = \frac{1}{2}\left(V_0 + \int_\Omega \phi\, dV\right), \quad (37)$$

where $V_0(\Omega)$ is the initial volume.

Likewise, a Lagrange multiplier for the volume $\lambda_V$ can be added in order to completely ensure the preservation of volume along the temporal evolution of the $\phi$ parameter as

$$\frac{\partial \phi}{\partial t} = M(\nabla^2 \mu_m + \lambda_V \frac{\delta V[\phi]}{\delta\phi}) = M(\nabla^2 \mu_m + \lambda_V), \quad (38)$$

where the Lagrange constraint for the volume $\dfrac{\delta V[\phi]}{\delta\phi} = 1$. This volume Lagrange multiplier $\lambda_V$ is also adopted from the penalty method [2]. In other previous studies [1, 5] the Lagrange multipliers are taken differently, as the Model-B dynamics conserve the volume of the system thus a volume Lagrange multiplier being unnecessary. However in this paper, at high flow velocities the volume is not perfectly conserved to the the small numerical errors adding up over time. Thus, we add this volume Lagrange multiplier.

**Deformable bodies in a 3-dimensional viscous flow: Vorticity-Stream vector formulation**

## RESULTS AND DISCUSSION

### Cahn Hilliard, a surface tension driven interface

A Canh-Hilliard interface, like an oil dropplet on water, is in general spherical and will not deform until very high flows. In Fig. 1 **D** we can see such deformations as we get some asymmetric shapes. This is the result of a droplet starting position being outside of the axis of the cylindrical channel. Results for droplets of oil when centered of the channel are, as expected, symmetrical and the shape tends towards a bullet shape with increasing velocity.

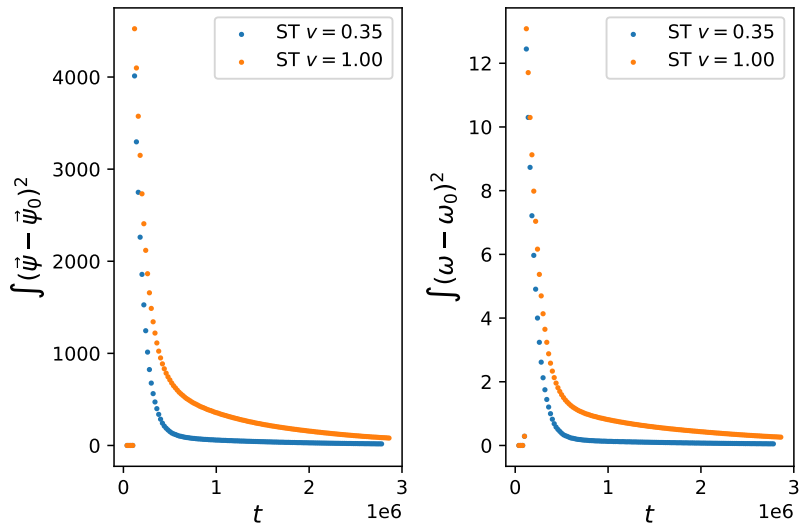

FIG. 1: For a straightforward surface tension interfacial energy the observables computed just give an exponential relaxation. The characteristic relaxation time increasing with the velocity of the Poiseuille flow.

### Membranes in Couette flow

The temporal evolution of bending energy in Figure 2 provides further evidence on the effect of how wall speed changes how much membranes deform and how long it takes. At the lower wall speed ($v_{wall} = 0.2$), the vesicle rapidly relaxes to a nearly constant bending energy, consistent with a stable equilibrium position. In contrast, at the higher wall speed ($v_{wall} = 0.8$), the bending energy displays extended transients and intermittent plateaus, reflecting a more complex migration pathway before stabilization. The final vesicle configurations (main Figure 4 **A,B**) clearly illustrate this contrast: the low-speed case preserves a relatively compact shape, whereas the high-speed case undergoes pronounced elongation under the stronger shear.

On lateral migration we have seen the effect of the viscosity contrast in 2D, which is to introduce a difference between the viscosity of the cell fluid and the channel fluid. This is represented with the viscosity contrast $\zeta$ which, as seen previously, takes an expression of

$$\zeta = \frac{\eta_{cell}}{\eta_{liquid}}. \tag{39}$$

We have only seen results for cells that have the same reduced volume as a red blood cell, giving it red blood cell shape when no flow is applied. The expression for the reduced volume for a cell of a given area value

$$V_r = \frac{V}{\frac{4}{3}\pi(\frac{A}{4\pi})^{3/2}}, \tag{40}$$

where $V$ is the membrane volume and $A$ the area. But in 2D we have also seen that by changing this reduced volume to more spherical values can change the results of the stationary position in a Couette flow like we can see in Figure 2.

**Deformable bodies in a 3-dimensional viscous flow: Vorticity-Stream vector formulation**

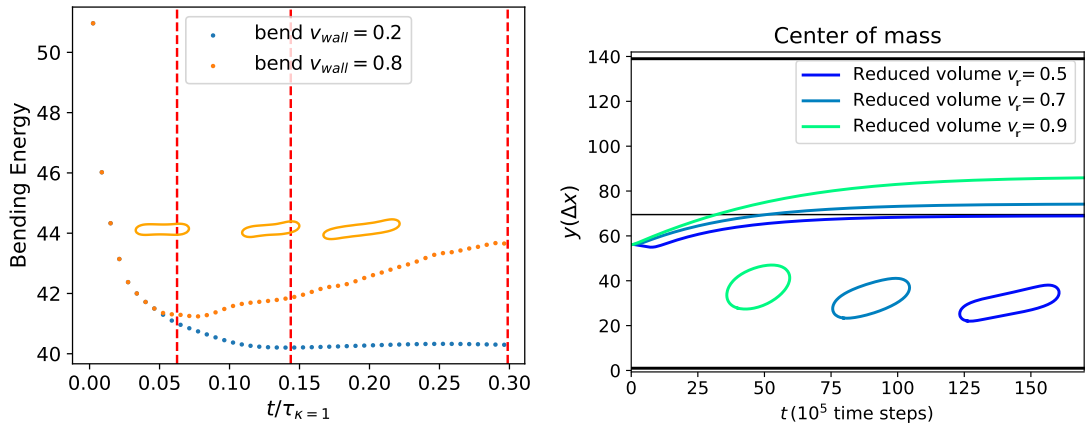

FIG. 2: Left: Bending energy over time of two simulations of membranes in a Couette flow. Right: Lateral migration effect in 2D changin with reduced volume.

* Electronic address: fdzgallen@gmail.com
[1] F. Campelo and A. Hernandez-Machado, *The European Physical Journal E*, 2006, **20**, 37–45.
[2] Q. Du, C. Liu and X. Wang, *Journal of Computational Physics*, 2004, **198**, 450–468.
[3] C. Peco, A. Rosolen and M. Arroyo, *Journal of Computational Physics*, 2013, **249**, 320–336.
[4] H. Goldstein, C. Poole and J. Safko, *Classical mechanics*, 2002.
[5] A. F. Gallen, J. R. Romero-Arias, R. A. Barrio and A. Hernandez-Machado, *Soft Matter*, 2023, **19**, 2908–2918.

 
\end{document}